\documentclass[twocolumn]{jpsj2} 
\title{Unconventional Superconductivity and Nearly Ferromagnetic Spin Fluctuations in Na$_x$CoO$_2$$\cdot~y$H$_2$O}

\author{
K. \textsc{Ishida}$^{1}$\thanks{E-mail address: kishida@scphys.kyoto-u.ac.jp}, 
Y. \textsc{Ihara}$^{1}$
Y. \textsc{Maeno}$^{1}$
C. \textsc{Michioka}$^{2}$
M. \textsc{Kato}$^{2}$
K. \textsc{Yoshimura}$^{2}$
K. \textsc{Takada}$^{3}$
T. \textsc{Sasaki}$^{3}$
H. \textsc{Sakurai}$^{4}$
and
E. \textsc{Takayama-Muromachi}$^{5}$}

\inst{$^{1}$Department of Physics, Graduate School of Science, Kyoto University, Kyoto 606-8502, Japan, \\
$^{2}$Department of Chemistry, Graduate School of Science, Kyoto University, Kyoto 606-8502, Japan \\
$^{3}$Advanced Materials Laboratory, National Institute for Materials Science, 1-1 Namiki, Tsukuba, Ibaraki, 305-0044, Japan\\
$^{4}$Superconducting Materials Center, National Institute for Materials Science, 1-1 Namiki, Tsukuba, Ibaraki 305-0044, Japan.}

\abst{Co NQR studies were performed in recently discovered superconductor Na$_x$CoO$_2$$\cdot y$H$_2$O to investigate physical properties in the superconducting (SC) and normal states. 
Two samples from the same Na$_x$CoO$_2$ were examined, SC bilayer-hydrate sample with $T_c \sim 4.7$ K and non-SC monolayer-hydrate sample. 
From the measurement of nuclear-spin lattice relaxation rate $1/T_1$ in the SC sample, it was found that the coherence peak is absent just below $T_c$ and that $1/T_1$ is proportional to temperature far below $T_c$. These results, which are in qualitatively agreement with the previous result by Fujimoto {\it et al.}, suggest strongly that unconventional superconductivity is realized in this compound. In the normal state, $1/T_1T$ of the SC sample shows gradual increase below 100K down to $T_c$, whereas $1/T_1T$ of the non-SC sample shows the Korringa behavior in this temperature range.
From the comparison between $1/T_1T$ and $\chi_{\rm bulk}$ in the SC sample, the increase of $1/T_1T$ is attributed to nearly ferromagnetic fluctuations. 
These remarkable findings suggest that the SC sample possesses nearly ferromagnetic fluctuations, which are possibly related with the unconventional superconductivity in this compound. The implication of this finding is discussed.}

\kword{superconductivity, hydrous sodium cobalt oxide, NQR, spin fluctuations}

\begin{document}
\maketitle

Superconductivity in Na$_x$CoO$_2$$\cdot y$H$_2$O ($x \sim 0.35$, $y \sim 1.3$) with the transition temperature $T_c \sim 5$ K was discovered quite recently by Takada {\it et al}\cite{Takada1}. 
Although $T_c$ is by one-order of magnitude smaller than that in cuprate superconductors, much attention have been paid for its unique crystal structure of the 2-dimensional (2-D) layer where superconductivity occurs. 
The CoO$_2$ forms a 2-D hexagonal layered structure, which is in contrast with the tetragonal structure of cuprates. 
Due to the vacancy of the Na atom, 0.65 holes are doped into the band insulating state of low-spin Co$^{3+}$ ($3d^6$ in $t_{2g}$ orbits), which is regarded alternatively as 0.35-electron doping state in the triangular lattice consisting of $S = 1/2$ of Co$^{4+}$. 
One may expect the unconventional superconductivity with magnetic frustrations originating from the triangular lattice.

For understanding superconducting (SC) properties, determination of the pairing symmetry is one of the most important issues.
For the purpose, nuclear magnetic resonance (NMR) and nuclear quadrupole resonance (NQR) measurements are suitable because they give crucial information about the orbital and spin states of the SC pairs from a microscopic point of view. 
In particular, NQR technique is powerful when some impurity phases are included in a sample, because NQR can spectroscopically detect only a concerned phase from the difference of the electric field gradient (EFG) in each phase. 
Furthermore, by using the NMR and NQR techniques, nuclear-spin lattice relaxation rate $1/T_1$ can be measured, which is related with low-energy spin dynamics in compounds.
  
Until now, there are three NMR and NQR reports on Na$_x$CoO$_2$$\cdot y$H$_2$O\cite{Waki,Kobayashi1,Fujimoto}. 
Two groups report that $1/T_1T$ is enhanced just below $T_{c}$\cite{Waki,Kobayashi1}. 
On the other hand, Fujimoto {\it et al.} show that $1/T_1$ decreases suddenly just below $T_c$ and follows the Korringa behavior ($T_1T$ = const. behavior) far below $T_c$\cite{Fujimoto}. 
Their observation on $1/T_1$ can be interpreted by the unconventional SC model with line-node gap. 
Thus the results reported so far contradict each other. 
To settle the controversy over the NMR and NQR results in the SC state, further $1/T_1$ studies are highly desired.

In this paper, we show our $1/T_1$ results measured independently. 
Our measurement was performed in the wider temperature range between 65mK and 200 K and on two samples with different character.
One of the samples is SC bilayer-hydrate Na$_x$CoO$_2$$\cdot y$H$_2$O with higher $T_c \sim$ 4.7 K than in the previous reports\cite{Kobayashi1,Fujimoto}. 
The other is the non-SC monolayer hydrate Na$_x$CoO$_2$$\cdot y$H$_2$O due to partial extraction of H$_2$O molecules between CoO$_2$ layers. 
Our $1/T_1$ result in the SC state is consistent with that by Fujimoto {\it et al.}, {\it i.e.} absence of the coherence peak just below $T_c$ and existence of the residual density of states far below $T_c$\cite{Fujimoto}. 
These are characteristic features of unconventional superconductivity. 
In addition, we found the low-energy spin-fluctuations in the SC sample, which is considered as nearly ferromagnetic fluctuations.
These results suggest that unconventional superconductivity appears in the metallic state with nearly ferromagnetic fluctuations, which play an important role for the occurrence of superconductivity.

We used powder sample for our NQR measurements, preparation of which was described in literatures\cite{Takada1,Takada2}.
SC transition was confirmed by dc-susceptibility measurement ($\chi_{\rm bulk}$).
The non-SC sample was obtained by storing the SC sample in the vacuum space at room temperature for three days.
From X-ray measurement, the $c$-axis lattice constant was found to 6.9 \AA, which corresponds to that in monolayer-hydrate sample\cite{Takada2}.
The bilayer-hydrate phase could not observed in the non-SC sample at all.
Since the non-SC sample was prepared from the SC sample, Na content is equivalent in two samples.
This is very important for considering the role of H$_2$O layer, since the superconductivity appears in quite narrow Na-concentration region\cite{Schaak}.

\begin{figure}[tb]
\begin{center}
\includegraphics[width=7cm]{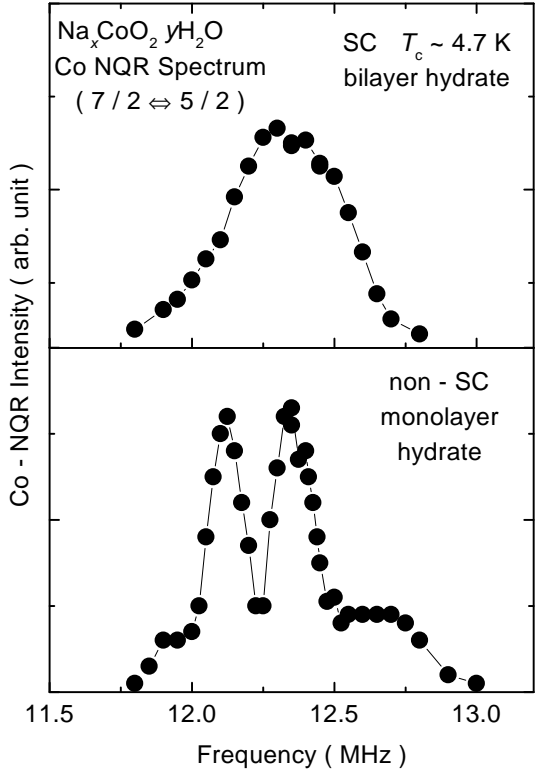}
\end{center}
\caption{Co-NQR spectra corresponding to $\pm5/2 \leftrightarrow \pm7/2$ transition in the SC bilayer-hydrate and non-SC monolayer-hydrate samples. The spectra were obtained by frequency-swept method.}
\label{f1}
\end{figure}

Figure 1 shows the Co-NQR spectra in the SC and non-SC samples, which were obtained by frequency-swept method.
The Co-NQR spectra originate from $^{57}$Co ($I$ = 7/2) nuclear-level transition between $\pm 5/2 \leftrightarrow \pm7/2$.
The SC sample shows a single peak at 12.35 MHz with the linewidth of 0.4 MHz.
The EFG frequency and the asymmetric parameter are consistent with previous reports\cite{Kobayashi1,Fujimoto}. 
On the other hand, the non-SC sample shows two sharp peaks at 12.1 and 12.35 MHz, together with other small satellite peaks at lower and higher frequency sides.
It seems that all Co sites are crystallographycally unique in the SC sample: the CoO$_2$ layer is sandwitched by two hydrate layers.
On the other hand, due to extraction of one hydrate layer, some Co sites in the CoO$_2$ layer of the non-SC sample have a different local symmetry from those of the SC sample, $e.g.$ the sharp peak at 12.1 MHz arises from the Co site, in which one of the hydrate molecules near Co is regularly replaced by Na, and the satellite peaks are from the Co sites, in which two or three hydrate molecules are replaced by Na.
For identification of Co-NQR peaks in the non-SC sample, further NQR studies are needed in different stoichiometric samples.  
 
$T_1$ was measured at 12.35 MHz in the SC sample and 12.1 and 12.35 MHz in the non-SC sample, respectively.
At these peaks, the recovery of the nuclear magnetization after saturation pulses can be fitted by the theoretical curve\cite{MacLaughlin} in whole temperature range except for very low temperature below 300 mK.

\begin{figure}[tb]
\begin{center}
\includegraphics[width=7cm]{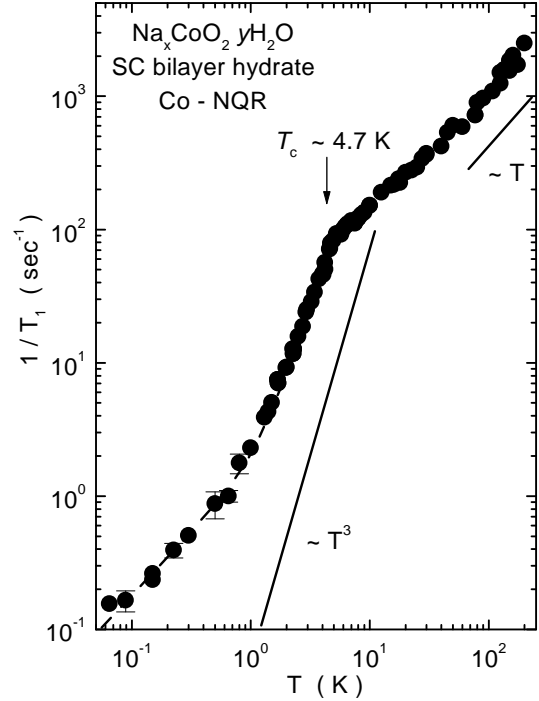}
\end{center}
\caption{Temperature dependence of $1/T_1$ of the SC bilayer-hydrate sample plotted in logarithmic scale. The dotted curve in the SC state is the calculation using the 2-D line-node ($\Delta(\phi) = \cos(2\phi)$ ) model with residual DOS induced by unitarity impurities. The appropriate fitting parameters are 2$\Delta / k_BT_c$ = 3.5 and $N_{res}/N_0 \sim 0.32$.}
\label{f2}
\end{figure}

First, we discuss temperature dependence of $1/T_1$ in the SC state.
Figure 2 shows temperature dependence of $1/T_1$ in logarithmic scale. 
$1/T_1$ shows sharp decrease below $T_c$ and crosses over to the Korringa behavior in low temperatures, which is qualitatively in good agreement with the result by Fujimoto {\it et al.}\cite{Fujimoto} 

\begin{figure}[tb]
\begin{center}
\includegraphics[width=7cm]{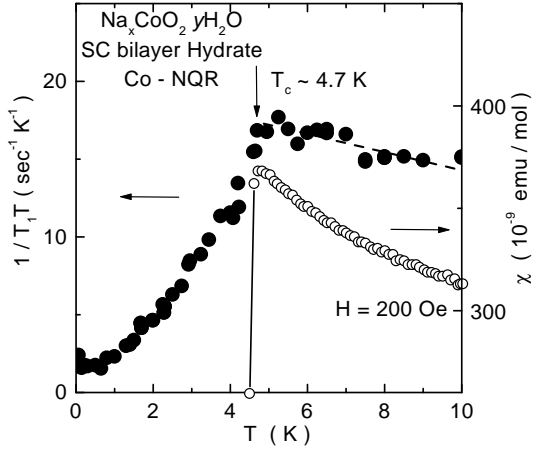}
\end{center}
\caption{Temperature dependence of $1/T_1T$ and $\chi_{\rm bulk}$ of the SC bilayer-hydrate sample below 10 K.}
\label{f3}
\end{figure}
To see the behavior around $T_c$ in detail, we show in Fig. 3 temperature dependence of $1/T_1$ divided by $T$, $1/T_1T$ of the SC sample below 10 K. 
For comparison, we also plot bulk susceptibility showing SC transition. 
As seen in the figure, $1/T_1T$ decreases just below $T_c$ determined by the susceptibility. 
It is obvious that the coherence peak is absent below $T_c$, which suggests that the superconductivity should be classified to an unconventional superconductor such as heavy-Fermion, cuprate, ruthenate, and organic superconductors. 

The whole temperature dependence in the SC state can be understood by the 2-D line-node model with $\Delta(\phi) = \Delta_0\cos(2\phi)$ incorporated with residual density of states ($N_{\rm res}$) ascribed to impurities and/or crystal imperfections\cite{IshidaZn}. 
The dotted lines in Fig.~2 is the calculation using the model with $2\Delta/k_BT_c$ = 3.5 and $N_{\rm res}/N_0 \sim 0.32$, where $N_0$ is the density of states at $T = T_c$. 
We found that $N_{\rm res} / N_0$ is smaller in the higher-$T_c$ samples than those  reported by Fujimoto {\it et al.}\cite{Fujimoto} ($N_{res} / N_0 \sim$ 0.65 and $T_c \sim 3.9$ K ). 
Similar tendency was already seen in unconventional superconductor Sr$_2$RuO$_4$\cite{IshidaSRO}. 
Based on the theoretical model by Hotta\cite{Hotta}, we can tentatively estimate $T_{c0}$ in a pure sample of this compound as 5.5 K.
Due to the presence of the residual density of states, we cannot rule out the possibility of the isotropic gap ascribed to $D$ + i$D$ state, but the most promising gap would be the line-node gap with residual density of states induced by unitarity impurities\cite{Bang}.    

\begin{figure}[tb]
\begin{center}
\includegraphics[width=7cm]{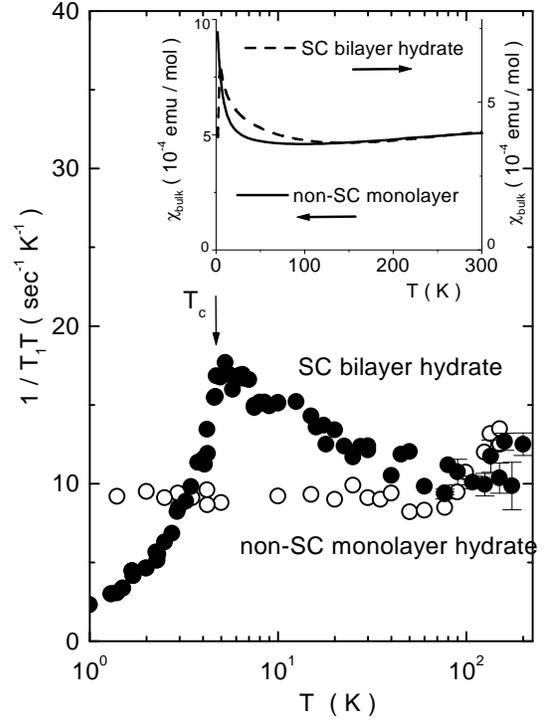}
\end{center}
\caption{Temperature dependence of $1/T_1T$ in SC bilayer-hydrate and non-SC monolayer-hydrate samples. The inset shows temperature dependence of $\chi_{\rm bulk}$. In the inset, $\chi_{\rm bulk}$ is normalized, so that the value of $\chi_{\rm bulk}$ at 300 K is identical in two samples.}
\label{f4}
\end{figure}

Next we move on to the normal-state properties in these compounds.
Temperature dependence of $1/T_1T$ of SC and non-SC samples, which was measured at 12.35 MHz, are shown in the main panel of Fig.~4, in which the scale of $T$ axis is in the logarithmic scale.
$1/T_1T$ was also measured at 12.1MHz in the non-SC sample, and the value of $1/T_1T$ is 8.2 $s^{-1}K^{-1}$ at 4.2 K, 10 \% smaller than $1/T_1T$ at 12.35 MHz, and the temperature dependence is the same as that at 12.35 MHz.
As seen in the figure, $1/T_1T$ of the non-SC sample shows the Korringa behavior from 100 K to 1.4 K, whereas $1/T_1T$ of the SC sample increases with decreasing temperature below 100K down to $T_c$.
The Korringa behavior is not seen at all just above $T_c$ in the SC sample.
It is obvious that the low-energy spin-fluctuations are present in the SC sample, which is not seen in the non-SC sample.

To understand the spin-fluctuation character, we compare the behavior of $1/T_1T$ with that of bulk susceptibility $\chi_{\rm bulk}$.
The inset of Fig.~4 shows the temperature dependence of $\chi_{\rm bulk}$ in two samples. 
The gradual increase of $\chi_{\rm bulk}$ in the SC sample is seen below 100 K, where $1/T_1T$ also increases. 
On the other hand, $\chi_{\rm bulk}$ in the non-SC sample shows a sharp increase below 50 K whereas $1/T_1T$ does not in this temperature range.
It seems that the increase of $\chi_{\rm bulk}$ in the SC and non-SC samples might be different in origin, $e.g.$ the sharp increase in the non-SC sample originates from local moments in some impurity phases, but the gradual increase in the SC sample is intrinsic behavior in the compound. 
This possibility is also suggested by the comparison between $1/T_1T$ and $\chi_{\rm bulk}$ in the SC sample as shown later.  

In general, $1/T_1T$ is related with low-energy part of the $q$-dependent dynamical susceptibility in compounds. 
From the comparison between $1/T_1T$ and $\chi_{\rm bulk}$ or Knight shift, we can have an important information about spin-fluctuation character, since the latter are related with the static susceptibility at $q = 0$, $\chi(0)$. 
If AFM correlations are dominant, which is the case in cuprate superconductors, dynamical susceptibility has peaks at the AFM wave vector $Q$ apart from $q = 0$, therefore $1/T_1T$ is mainly determined by staggered susceptibility, $\chi(Q)$\cite{Kitaoka}. 
In most cuprate superconductors, $1/T_1T$ and $\chi_{\rm bulk}$ show different behavior, {\it i.e.}, $1/T_1T$ is enhanced whereas $\chi_{\rm bulk}$ decreases with deceasing temperature due to the development of AFM correlations\cite{Takigawa}. 
Such information played a crucial role in identifying the existence of AFM spin fluctuations in underdoped cuprate superconductor. 
On the contrary, when FM correlations are dominant, the dynamical susceptibility shows a peak at $q$ = 0, thus $1/T_1T$ is dominant by $\chi(0)$ component. 
In the SCR theory which describes the magnetic properties in weakly or nearly FM metallic state successfully, $1/T_1T$ was suggested to be proportional to $\chi(0)$\cite{SCR}. 
This relation was confirmed experimentally in nearly FM metallic compounds such as Pd\cite{TakigawaPd} and YCo$_2$\cite{Yoshimura}. 
The nearly FM fluctuations originate from the high density of states of $d$-electron at the Fermi level.

\begin{figure}[tb]
\begin{center}
\includegraphics[width=7cm]{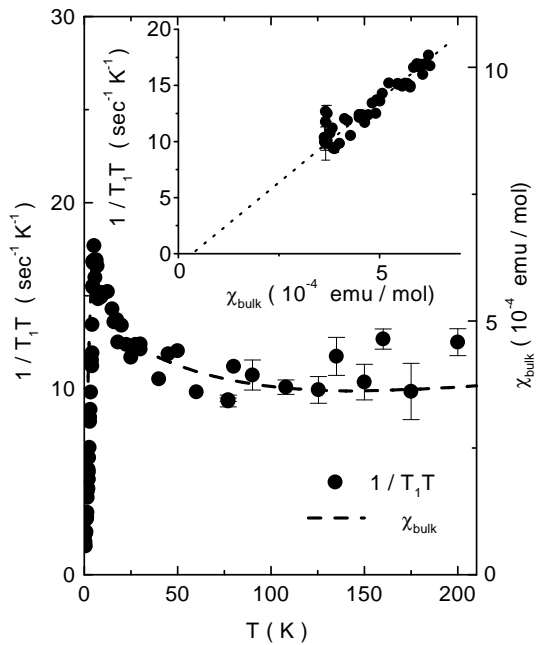}
\end{center}
\caption{Temperature dependence of $1/T_1T$ and $\chi_{\rm bulk}$ up to 200 K. The inset shows the plot of $1/T_1T$ against $\chi_{\rm bulk}$.}
\label{f5}
\end{figure}

Figure 5 shows temperature dependence of $1/T_1T$ in the normal state up to 200 K, together with bulk susceptibility $\chi_{\rm bulk}$ for comparison. 
It was found that $1/T_1T$ shows a good linear relation with $\chi_{\rm bulk}$, as seen in the inset of Fig.~5\cite{comment}. 
The good linear scaling between $1/T_1T$ and $\chi_{\rm bulk}$ strongly suggests that upturn of $\chi_{\rm bulk}$ in the SC sample below 100 K is intrinsic behavior in the compound, not from impurity phases. 
The linear relation in the inset of Fig.~5 also suggests that the spin fluctuations seen in the SC sample is nearly ferromagnetic ones, and that superconductivity would appear in the nearly FM metallic state, which is a quite contrast situation where cuprate superconductivity appears. 
The nearly FM state is also suggested by the large Wilson ratio (3.33), where the values of $\chi_{\rm bulk}$ and $\gamma$ of specific-heat measurement just above $T_c$\cite{Ueland} are adopted for the estimation.
 
Until now, it was found that most unconventional superconductors appearing in the paramagnetic state have strong AFM spin fluctuations. 
Even in spin-triplet superconductors reported so far, {\it i.e.}, UPt$_3$\cite{UPt3} and Sr$_2$RuO$_4$\cite{Maeno}, dominant spin fluctuations are not FM but AFM like. 
Recently unconventional superconductivity was discovered far below FM transition, which is considered as spin-triplet superconductors in the FM state\cite{UGe2,URhGe}.  
However, although the existence of FM fluctuations above $T_c$ has not been identified in the FM superconductors due to difficulty of experiments, it is likely that FM fluctuations are mostly quenched deep into the FM ordered phase. 
There has been no report about superconductivity appearing in the predominance of FM fluctuations. 
Thus, as far as we know, Na$_x$CoO$_2$$\cdot y$H$_2$O reported here is the first example that the superconductivity appears in the nearly FM metallic state. 
In such a situation, spin-triplet superconductivity in analogy with superfluidity of $^3$He would be the most promising state within possible SC pairing states.
To identify the spin state of SC pairs, Knight-shift measurement in the SC state is most crucial. 
Quite recently, Waki {\it et al.} reported that the Co-NMR spectrum in the aligned SC sample is unchanged on passing through $T_c$\cite{Waki}, whereas Kobayashi {\it et al.} reported the decrease of the Knight shift in the non-aligned SC sample\cite{Kobayashi2}. The Knight shift behavior in the SC state is now controversial issue.  
The result of the former group suggests a spin-triplet superconducting 
state which is consistent with our NQR results.

In conclusion, we show by Co-NMR study that unconventional superconductivity is realized in Na$_x$CoO$_2$$\cdot y$H$_2$O. 
The promising SC-gap would be line-node one with residual density of states induced by unitarity impurities. 
We found the low-energy spin-fluctuations present only in the SC sample, which are considered as nearly ferromagnetic fluctuations from the comparison between $1/T_1T$ and $\chi_{\rm bulk}$.
We suggest that Na$_x$CoO$_2$$\cdot y$H$_2$O is a new type of superconductivity, in which nearly FM spin fluctuations play the primary role for the mechanism of superconductivity.

We thank H.Yaguchi for experimental support, and H. Ikeda, and K. Yamada for valuable discussions.
One of the authors (K.I.) thanks W. Higemoto, and R. Kadono for valuable information. 
This work has been partially supported by CREST of Japan Science and Technology Corporation (JST) and the 21COE Research in Grant-in-Aid for Scientific Research from the Ministry of Education, Culture, Sport, Science and Technology of Japan (MECSST), and by the Grants-in-Aid for Scientific Research from Japan Society for Promotion of Science (JSPS), and MECSST.


\begin{thebibliography}{99}
\bibitem{comment}
The intercept of the $x$ axis gives the orbital Van-Vleck susceptibility in the compound, which is estimated as 2 $\times 10^{-5}$ emu / mol. This suggests that the dominant contribution is from the spin susceptibility of the Co-$d$ electrons
\end{thebibliography}
\end{document}